\documentclass[12pt]{article}

\DeclareMathAlphabet{\scr}{U}{rsfs}{m}{n}
\usepackage{latexsym}
\usepackage[mathscr]{eucal}
\usepackage{amsfonts}
\usepackage{amscd}
\usepackage{cite}
\usepackage{array}   
\usepackage{amssymb}
\usepackage{colordvi}
\usepackage[centertags]{amsmath}
\usepackage{enumerate}
\usepackage{graphicx}
\usepackage{booktabs}
\usepackage{theorem}
\usepackage[footnotesize]{caption}
\usepackage{subcaption}

\newcommand{\newc}{\newcommand}
\newc{\be}{\begin{equation}}
\newc{\ee}{\end{equation}}
\newc{\bea}{\begin{eqnarray}}
\newc{\eea}{\end{eqnarray}}
\newc{\ol}{\overline}
\newc{\wt}{\widetilde}
\newc{\bs}{\boldsymbol}
\newc{\m}{\mathcal}
\newc{\la}{\langle}
\newc{\ra}{\rangle}

\usepackage[latin1]{inputenc}

\usepackage{amssymb}
\usepackage{graphicx}
\usepackage{epsfig}
\usepackage{psfrag}
\usepackage{latexsym}
\usepackage{indentfirst}
\usepackage{fancyhdr}
\usepackage{amssymb}
\usepackage{amsmath}
\usepackage{amsfonts}
\usepackage{colordvi}
\usepackage{pifont}
\usepackage{color}
\usepackage{graphicx}
\usepackage{url}
\usepackage{array}
\usepackage{rotating}

\makeatletter
\def\Ddots{\mathinner{\mkern1mu\raise\p@
\vbox{\kern7\p@\hbox{.}}\mkern2mu
\raise4\p@\hbox{.}\mkern2mu\raise7\p@\hbox{.}\mkern1mu}}
\makeatother

\def\beq{\begin{equation}}
\def\eeq{\end{equation}}
\def\be{\begin{equation}}
\def\ee{\end{equation}}
\def\bea{\begin{eqnarray}}
\def\eea{\end{eqnarray}}
\def\wt{\widetilde}
\def\ol{\overline}

\begin{document}
%\vspace{0.5in}
\title{\vfill ~\\[-30mm] \hfill ~\\[-30mm]
      % \hfill\mbox{\small SHEP-11-36}\\[-3.5mm]
%\hfill\mbox{\small IPPP-11-83}\\[-3.5mm]
%\hfill\mbox{\small DCPT-11-166}\\[13mm]
       \textbf{Lepton Mixing Predictions including Majorana Phases from $\Delta(6n^2)$ Flavour Symmetry and Generalised CP\\[4mm]}}

\date{}

\date{}
\author{
Stephen~F.~King$^{1\,}$\footnote{E-mail: \texttt{king@soton.ac.uk}}~~and~
Thomas Neder$^{1\,}$\footnote{E-mail: \texttt{T.Neder@soton.ac.uk}}
\\[9mm]
{\small\it
$^1$School of Physics and Astronomy,
University of Southampton,}\\
{\small\it Southampton, SO17 1BJ, U.K.}\\[3mm]
}

\maketitle

\begin{abstract}
\noindent  
Generalised CP transformations are the only known framework which allows to predict Majorana phases in a flavour model purely from symmetry. For the first time generalised CP transformations are investigated for an infinite series of finite groups, $\Delta(6n^2)=(Z_n\times Z_n)\rtimes S_3$. In direct models the mixing angles and Dirac CP phase are solely predicted from symmetry. $\Delta(6n^2)$ flavour symmetry provides many examples of viable predictions for mixing angles. For all groups the mixing matrix has a trimaximal middle column and the Dirac CP phase is 0 or $\pi$. The Majorana phases are predicted from residual flavour and CP symmetries where $\alpha_{21}$ can take several discrete values for each $n$ and the Majorana phase $\alpha_{31}$ is a multiple of $\pi$. We discuss constraints on the groups and CP transformations from measurements of the neutrino mixing angles and from neutrinoless double-beta decay and find that predictions for mixing angles and all phases are accessible to experiments in the near future.
 \end{abstract}
\thispagestyle{empty}
\vfill
\newpage
\setcounter{page}{1}

%%%%%%%%%%%%%%%%%%%%%%%%%%%%%%%%%%%%%%%%%
\section{Introduction}
The question of the origin of neutrino masses and mixing parameters is of fundamental importance. One approach are so-called direct models of neutrino masses \cite{King:2013eh} where a discrete non-Abelian family symmetry group is broken to a $Z_2\times Z_2$ group in the Neutrino sector,
and a $Z_3$ subgroup in the charged lepton sector. In such a model the lepton mixing angles and the lepton Dirac CP phase are completely fixed by symmetry.

Recently such direct models have been analysed with the help of the group database GAP \cite{d150lam,Holthausen:2012wt}. The only flavour groups that can produce viable mixing parameters in a direct model belong to the group series $\Delta(6n^2)$  or are subgroups of such groups. The group theory of $\Delta(6n^2)$ groups has been analysed in \cite{Escobar:2008vc}. The consequences for neutrino mixing from a $\Delta(6n^2)$ flavour symmetry in direct models have been studied in detail in \cite{KingVNA} for arbitrary even $n$. Some examples of $\Delta(6n^2)$ groups or subgroups have previously been studied in \cite{TooropJN,deAdelhartTooropRE, DingXX, KingIN, LamNG, KingAP,VarzielasSS, KrishnanSB,HolthausenVBA}.

In the Standard Model, violation of CP occurs in the flavour sector. Promoting CP to a symmetry at high energies which is then broken allows to impose further constraints on mass matrices of charged leptons and Majorana neutrinos. In this case the interplay between CP and flavour symmetries has to be carefully discussed\cite{ChenTPA,GirardiSZA,DingNSA,FeruglioCW,FeruglioHIA,NishiJQA,LuhnLKN,GrimusZI,DingBPA,DingHPA,HolthausenDK,KuchimanchiXB,KuchimanchiXS,KuchimanchiTE,BoucennaQB}. For direct models, especially with a flavour group from $\Delta(6n^2)$, CP symmetries have not been studied in detail yet.

In this paper we examine a class of generalised CP (gCP) transformations consistent with $\Delta(6n^2)$ groups for arbitrary $n$. We will start by defining flavour and generalised CP transformations and stating their effect on mass matrices. In the following section we review and develop the general theory of gCP transformations in the presence of flavour symmetries in a general context. Afterwards we specialise on direct models with $\Delta(6n^2)$ as a flavour group, where we compute the lepton mixing matrix including Majorana phases for arbitrary even $n$ for all possible breaking patterns of the flavour group and of gCP. Here we also analyse the constraints from measurements of the mixing angles and from neutrinoless double-beta-decay on these models. In the last section we conclude.

\section{Generalised CP Transformations, Flavour Symmetries, Automorphisms and the Character Table}
In this section we review the interplay between flavour symmetries and CP symmetries which has especially been discussed in \cite{ChenTPA,FeruglioCW,NishiJQA,GrimusZI,HolthausenDK} and use general arguments to show that for a class of groups $G$, of which 
$G=\Delta(6n^2)$ is an example, physical CP transformations correspond to $X_r\in e^{i\alpha} G$ with $\alpha$ a real number.

\subsection{Generalised CP transformations and flavour symmetries}
Consider a theory where generations of fermions are assigned to multiplets of representations $r$ of a flavour group $G$ and that is invariant under transformations of the multiplets $\varphi_r$ under the group $G$
\begin{equation}
\varphi_r\mapsto \rho_r(g)\varphi_r
\end{equation}
where $\rho_r(g)$ is the representation matrix for $g\in G$ in the representation $r$.

Further consider the group $G$ being broken to a Klein subgroup $G_\nu\simeq Z_2\times Z_2$ subgroup in the neutrino sector and an abelian subgroup $G_e\simeq Z_m$ with $m>2$ in the charged lepton sector. If these subgroups remain unbroken at all energies, in the low-energy-limit constraints on the mass matrices of charged leptons and neutrinos are imposed. Left-handed doublets transform under the same representation $r$. The charged lepton mass matrix $M^e$ has to fulfill 
\begin{equation}
\rho_r(g)^\dagger M^e(M^e)^\dagger \rho_r(g)=M^e(M^e)^\dagger
\end{equation}
with $\rho_r(g)$ being the representation matrix of $g\in G_e$ in the representation $r$. The Majorana neutrino mass matrix is constrained by
\begin{equation}
\rho_r(g)^T M^\nu \rho_r(g)=M^\nu
\end{equation}
with $g\in K_\nu$.

Define generalised CP (gCP) by
\begin{equation}
\varphi_r\mapsto X_r(\varphi_{r}^\ast(x^P)) 
\end{equation}
where $r$ is the representation of $G$ according to which $\varphi_r$ transforms. \footnote{Other Authors consider transformations of the type $\varphi_r\mapsto \varphi_{r'}^\ast$ where $r,r'$ can be different. In \cite{ChenTPA} has been shown that only gCP transformations where $r=r'$ actually make observables (e.g. particle decays) conserve CP.} $X_r$ is a unitary matrix. We need to find all matrices $X_r$ that are ``allowed'' in coexistence with a flavour group $G$. The aforesaid will be made a more precise statement in the following section, where the conditions for the existence of gCP transformations as well as their properties will be discussed. 

If the theory at the low-energy end is invariant under residual gCP transformations with matrices $X_r^e$ for charged leptons and $X_r^\nu$ for neutrinos then the mass matrices will be constrained by
\begin{equation}
X_r^{e\dagger} M^e (M^e)^\dagger X_r^e = (M^e)^\ast (M^e)^T
\end{equation} 
for charged leptons and by
\begin{equation}
X_r^{\nu T} M^\nu X_r^\nu = (M^\nu)^\ast
\end{equation}
for Majorana neutrinos.

If $X_r^\nu \in G_\nu$ ($X_r^e\in G_e$), no new constraints on the neutrino (charged lepton) mass matrix follow but it being real. With $g,h\in (Z_2\times Z_2)$ from $\rho_r(g)X_r\rho_r(h)$ only the same constraints as for $X_r$ follow for the mass matrix. This means only $X_r$ that are not in $(Z_2\times Z_2)$ allow for a mass matrix that is not real and at the same time impose new constraints on it. 

\subsection{The consistency equation}

We would like to know which transformations of the type
\begin{equation}
\varphi_r \mapsto X_r \varphi_r^\ast(x^P)
\label{gCP2}
\end{equation}
can be applied to the theory without destroying the invariance under $G$, i.e. which matrices $X_r$ can appear in Eq.(\ref{gCP2}) that preserve symmetry under $G$? 
Consider performing a gCP transformation followed by a flavour transformation followed by the inverse gCP transformation. From invariance of the theory under $G$ follows that the matrix $X_r$ is allowed in a gCP transformation if for every $g\in G$ there is a $g'\in G$ such that
\begin{equation}
X_r \rho_r^\ast(g)X_r^\dagger=\rho_r(g').
\label{consistency}
\end{equation}
Eq. (\ref{consistency}) is called the consistency equation and an $X_r$ that fulfills it is called consistent with $G$.

If $r$ is a faithful representation, which is equivalent to saying that $\rho_r$ is injective, one can define a bijective mapping $u_X:G\rightarrow G$ between the elements of the group:
\begin{equation}
u_X(g):=\rho_r^{-1}(X_r\rho_r^\ast(g)X_r^\dagger).
\label{ConsistencyEq}
\end{equation}
(One can drop the index $r$ on $u_{X_r}$ because for all faithfull irreps the mapping generated by Eq.(\ref{ConsistencyEq}) will be the same).
For faithful representations $r$, $u_X(g)$ is an automorphism of the group $G$. 
% This follows immediately if $\rho_r(g)$ is a real matrix, for complex $\rho_r(g)$ one has to deal with complex conjugation. As we are only considering unitary representations of $G$,
% \begin{equation}
% \rho^\dagger_r(g)=\rho(g^{-1}).
% \end{equation}
% For each irrep $r$ we can define a unitary matrix $w_r$ which implements matrix transposition on that irrep with which follows
% \begin{equation}
% \rho^\ast_r(g)=\rho_r^T(g^{-1})=w_r\rho_r(g^{-1})w_r^\dagger.
% \end{equation}
% For a general $\rho_r(g)$ the most general matrix $w_r$ is given by
% \begin{equation}
% w_r=e^{i \alpha_r}\begin{pmatrix}
%       & &1\\
%       &\Ddots&\\
%       1& &
%     \end{pmatrix}
% \end{equation}
% with a real number $\alpha_r$. The mapping produced by a matrix $X_r$ is given by
% \begin{equation}
% u_X(g)=\rho_r^{-1}(X_r w_r \rho_r(g^{-1}) w_r^\dagger X_r^\dagger)
% \label{u_X}
% \end{equation}
% whose being an automorphism can be checked. We will return to Eq. (\ref{u_X}) later and continue at this point by stating several properties of automorphisms which will be needed for analysing the function Eq.(\ref{u_X}).

\subsection{Inner and outer automorphisms}
Group automorphisms come in two kinds: Inner and outer automorphisms. Inner automorphisms $Inn(G)$ are such automorphisms $u:G\rightarrow G$ where for all $g\in G$ one single group element $h_u$ exists such that
\begin{equation}
u(g)=h_u^{-1}g h_u.
\label{inner}
\end{equation}
All inner automorphisms are given by $Inn(G)=G/Z(G)$, where $Z(G)$ is the center of G, i.e. all elements of $G$ that commute with every other group element. Outer automorphisms $Out(G)$ are all automorphisms that are not inner.

An inner automorphism will map each element into its original conjugacy class. An outer automorphism however is not inner which means that there is at least one $g'\in G$ for which with all $h\in G$ $u(g)\neq h^{-1}g'h$ (compare with the definition of inner automorphisms before Eq. (\ref{inner})), i.e. there is at least one $g'\in G$ which is not mapped back into its original conjugacy class. Also if $g$ is in the class $C_k$ and it is mapped onto $u(g)$ which is in the class $C_l$, every element in $C_k$ is mapped on an element in $C_l$ by $u$.

This proves also that an automorphism that maps each element back into its original conjugacy class is inner, as well that an automorphism that maps elements from at least two conjugacy classes on each other is outer.
\footnote{
An outer automorphism $u$ also generates mappings between different representations of $G$. For two representations $\rho_r$ and $\rho_s$ define
\begin{equation}
u_{sr}=\rho_s\circ u \circ \rho_r^{-1}
\end{equation}
with which follows
\begin{equation}
(u_{sr}\circ \rho_r)(g)=\rho_s(u(g)).
\end{equation}

The outer automorphism $u$ acting inside the group thus interchanges columns of the character table while when acting between representations via $u_{sr}$ interchanges rows of the character table. We call a symmetry of the character table 
\begin{equation}
\chi_{jk}=\text{tr} \rho_j(g_k),\text{ }g_k\in C_k 
\end{equation}
any transformation of the type 
\begin{equation}
\chi_{jk}\mapsto P_{ij}\chi_{kl}Q_{kl}
\end{equation}
with permutation matrices $P$ and $Q$ that leaves $\chi$ invariant, i.e. 
\begin{equation}
P_{ij}\chi_{kl}Q_{kl}=\chi_{ij}
\end{equation}
and where only classes of the same size and element-order are interchanged, i.e $|C_l|=|C_j|$ and $\text{ord}g_l=\text{ord}g_j$ for $g_l\in C_l$ and $g_j\in C_j$.
An outer automorphism will always generate a non-trivial symmetry of the character table, just as a symmetry of the character table always gives rise to an outer automorphism: Define the automorphism by the action on the conjugacy classes, a corresponding permutation of the representations is always given by any outer automorphism via $u_{sr}$.}

We will now return to the automorphism $u_X$ (\ref{ConsistencyEq}) that is induced by the consistency equation (\ref{consistency}). If $\rho_r(g)$ is real and $X_r\in G$ then $u_X$ will be an inner automorphism. This is also true if $X_r\in e^{i \alpha}G$. 

If on the other hand $u$ is an outer automorphism it follows that a matrix $X_r$ that could mediate $u$ \'{a} la Eq. (\ref{ConsistencyEq}) is not in $e^{i\alpha}G$ (if it exists).

One could ask now if there can be a matrix $\tilde{X}_r$ that is not in $e^{i\alpha}G$ for that $u_{\tilde{X}}$ only connects elements within the same conjugacy class, i.e. that generates an inner automorphism? As for an inner automorphism $u$ there always is a single $h_u\in G$ such that the automorphism is given by $u(g)=h_u^{-1}gh_u$ it follows that
\begin{equation}
\tilde{X}_r\rho_r^\ast(g_k)\tilde{X}_r^\dagger=\rho_r(h_u)\rho_r(g_k)\rho_r(h_u^{-1}).
\end{equation}

For a real matrix $\rho_r(g)$ multiplying by  $\tilde{X}_r$ from the right and by $\rho_r(h_u^{-1})$ from the left yields
\begin{equation}
\rho_r(h_u^{-1})\tilde{X}_r\rho_r(g_k)=\rho_r(g_k)\rho(h_u^{-1})\tilde{X}_r.
\end{equation}
As $g_k$ can be every element of $G$, $\rho_r(h_u^{-1})\tilde{X}_r$ commutes with every group element. One can now apply Schur's Lemma \footnote{To be precise one uses the second part of Schur's Lemma which states that an operator that in some representation commutes with every group element is proportional to the identity.} to find that 
\begin{equation}
\tilde{X}_r=\lambda \rho_r(h_u)
\end{equation}
where $|\lambda|=1$ to keep $\tilde{X}_r$ unitary. As $\tilde{X}_r$ was supposed to not be in $e^{i\alpha}G$ this is in contradiction to the assumptions. For real $\rho_r(g)$ this proves that inner automorphisms correspond to $X\in e^{i\alpha}G$. For real representations, there is always a basis where this is the case, i.e where $\rho_r(g)$ is real for every $g\in G$.

If $\rho_r(g)$ is complex one has to deal with complex conjugation: Assume there is a matrix $w_r$ such that by applying complex conjugation and this matrix on an element of $G$, the element is mapped into the class of its inverse, $C(g^{-1})$:
\begin{equation}
\rho_r(g)\mapsto w_r^\dagger \rho_r(g)^\ast w_r \in C(g^{-1}).
\end{equation}
This can be thought of as an automorphism mapping $g\mapsto g^{-1}$ followed by an automorphism that maps $g^{-1}$ onto another element in the same class. As in the second step every element is sent into the original class, this second mapping is an inner automorphism and therefore by definition a single group element $h$ exists which inverts this step such that 
\begin{equation}
\rho_r(h)^\dagger(w_r^\dagger \rho_r(g)^\ast w_r)\rho_r(h)=g^{-1}.
\end{equation}
For this reason we assume in the following that the matrix $w_r$ maps  elements directly onto their inverses.
Using this, the general mapping induced by the consistency equation is given by: 
\begin{equation}
u_X(g)=\rho_r^{-1}(X_rw_r\rho_r(g^{-1}) w_r^\dagger X_r^\dagger)
\end{equation}
This mapping can be seen as an automorphism mapping $g$ on $g^{-1}$ followed by an automorphism given by $X_rw_r$:
\begin{equation}
u_X(g)=u_{Xw}(g^{-1}).
\end{equation}
If both $w_r$ and $X_r$ are contained in $e^{i\alpha}G$, $u_{X}$ will map $g$ in the same conjugacy class as $g^{-1}$. 
For $\Delta(6n^2)$, $w_r=\rho_r(b)$ maps elements into the class of the inverse and is contained in the group. We will not consider $w_r\notin G$ further.

Analogous to real irreps above one can now ask if there can be matrices $\tilde{X}_r$ that are not in $e^{i\alpha}G$ but that with $w_r\in e^{i\alpha} G$ will map $g$ in the conjugacy class of $g^{-1}$? This would be equivalent to $u_{\tilde{X}w}$ being an inner automorphism which would mean that for each group element $g\in G$ there is a single $h_u\in G$ such that
\begin{equation}
\rho_r(h_u)\rho_r(g^{-1})\rho_r(h_u^{-1})=X_r w_r \rho_r(g^{-1})w_r^\dagger X_r^\dagger.
\end{equation}
Again we can use Schur's Lemma and find there is $\lambda\in \mathbb{C}\setminus\{0\}$ such that
\begin{equation}
X_r=\lambda \rho_r(h_u)w_r^\dagger
\end{equation}
with $|\lambda|=1$ to make $X_r$ unitary. This contradicts $X_r \notin e^{i \alpha} G$. 
We have proved now that if $w_r\in e^{i \alpha}G$ then if and only if $X\in e^{i \alpha}G$ $u_X(g)$ will be in the conjugacy class of $g^{-1}$. In \cite{ChenTPA} the authors show that only gCP transformations that map elements into the class of its inverse element make observables conserve CP. We have proved here that such transformations are given by $X_r\in e^{i\alpha} G$.
\footnote{
We would now be able to find all $X_r\notin e^{i \alpha}G$ by reading off all automorphisms from the symmetries of the character table that do not map the class of $g$ on the class of $g^{-1}$. (This would often contain the identity transformation on the character table.)} In the following we will specialise $G$ to be $\Delta(6n^2)$.

\section{$\text{gCP}$ Symmetries and $\Delta(6n^2)$ groups}

In this section we consider gCP transformations where $X\in e^{i \alpha}G$ for $G=\Delta(6n^2)$. First we derive the gCP transformations that are consistent with $G_\nu=Z_2\times Z_2$ and $G_2=Z_3$. Afterwards we state the constrained mass matrices and the lepton mixing matrix. After this we discuss constraints from measurements of lepton mixing angles and from neutrinoless double-beta decay for arbitrary $n$.
 
If we want to break the flavour symmetry to  $G_\nu=Z_2\times Z_2$ and $G_e=Z_3$ subgroups, the residual flavour and residual gCP transformations are not independent, as they still have to fulfill the consistency equation. If e.g. in one sector $\rho_r(g)$ and $X_r$ are unbroken, then also $X_r \rho_r(g)^\ast X_r^\dagger$ must be unbroken. 
Thus the allowed residual gCP transformations have to map elements from the Klein group in consideration into said Klein group. 

The Klein subgroups of $\Delta(6n^2)$ are given by \cite{KingVNA} 
\begin{eqnarray}\label{klein1}
\{1,c^{n/2},d^{n/2},c^{n/2}d^{n/2}\},\\\label{klein2} \{1,c^{n/2},abc^\gamma,abc^{\gamma+n/2}\},\\\label{klein3}
 \{1,d^{n/2},a^2bd^\delta,a^2bd^{\delta+n/2}\}, \\\label{klein4} \{1,c^{n/2}d^{n/2},bc^\epsilon d^\epsilon, bc^{\epsilon-n/2}d^{\epsilon-n/2}\},
\end{eqnarray}
where $\gamma, \delta, \epsilon=1,\ldots,n/2$. 
The group Eq. (\ref{klein1}) will produce a mixing matrix with $|V_{ij}|=1/\sqrt{3}$, we will not consider it further.
The bottom three Klein subgroups will generate the same mixing matrix, thus it is sufficient to only consider the mixing matrices generated by group Eq.(\ref{klein2}).
The allowed matrices $X_r$ in the low-energy-limit have to be contained in $e^{i\alpha}G_\varphi$. A matrix $X_r$ is allowed if for a Klein subgroup $K$ holds that for each $g\in K$ also $u(g)\in K$.
For said Klein subgroup $K=\{1,c^{n/2},abc^\gamma,abc^{\gamma+n/2}\}$ one finds that the allowed matrices $X\in e^{i\alpha}G$ are given by the representation matrices for 
\begin{equation}
X_r=\rho_r(e^{i\alpha}c^x d^{2x+2\gamma}),\rho_r(e^{i\alpha}c^xd^{2\gamma+2x+n/2}),\rho_r(e^{i\alpha}abc^xd^{2x}),\rho_r(e^{i\alpha}abc^xd^{2x+n/2})
\label{Klein1gCP}
\end{equation}
with $\alpha\in \mathbb{R}$ and $x=0,\ldots,n-1$.

Without loss of generality, left-handed doublets $(\nu_L,e_L)^T$ are assigned to the representation $3_2^1$ (c.f.\cite{KingVNA}). Invariance of the mass matrix under the Klein subgroup in consideration plus invariance under one of the transformations from Eq. (\ref{Klein1gCP}) constrains the Majorana neutrino mass matrix to

\begin{equation}
M_\nu=\begin{pmatrix}
|m_{22}|e^{2i\pi\frac{\gamma}{n}}e^{i\varphi_1}&|m_{21}|e^{i\varphi_1}&0\\
|m_{21}|e^{i\varphi_1}&|m_{22}|e^{-2i\pi\frac{\gamma}{n}}e^{i\varphi_1}&0\\
0&0&|m_{33}|e^{i\varphi_3}

      \end{pmatrix}
      \label{mnu}
\end{equation}
where the values of $\varphi_1$ and $\varphi_3$ can be found in table (\ref{phitable}). In principle, several gCP transformations can remain unbroken. However, the phases $\varphi_1,\varphi_3$ are already fixed by one single unbroken transformation. Leaving a second gCP transformation unbroken with incompatible constraints on the phase $\varphi_i$ will force the corresponding mass parameters $|m_{..}|$ to be zero. The masses of neutrinos are $|m_{33}|$ and $||m_{21}|\pm|m_{22}||$. Thus $|m_{21}|=0$ or $|m_{22}|=0$ will result in a pair of degenerate neutrino states. It is not possible to have $|m_{33}|=0$ without $|m_{21}|=0$ or $|m_{22}|=0$. Leaving a second gCP transformation unbroken is never physically viable.
\begin{center}
\begin{table}[h]
  \begin{tabular}{| l || l  |l|}
    \hline
    $X_r$ & $\varphi_1$ &$\varphi_3$\\
    \hline
    \hline
    $\rho_r(e^{i\alpha}c^x d^{2x+2\gamma})$ & $-\alpha-2\pi(\gamma+x)/n$ &  $-\alpha +4\pi(\gamma+x)/n$ \\ \hline
    $\rho_r(e^{i\alpha}c^xd^{2\gamma+2x+n/2})$ & $-\alpha-\pi/2-2\pi(\gamma+x)/n$ &  $-\alpha+\pi+4\pi(\gamma+x)/n$ \\ \hline
    $\rho_r(e^{i\alpha}abc^xd^{2x})$ & $-\alpha-2\pi x/n$ &  $-\alpha+4\pi x/n$\\ \hline    
    $\rho_r(e^{i\alpha}abc^xd^{2x+n/2})$&$-\alpha-\pi/2-2\pi x/n$&$-\alpha+\pi+4\pi x/n$\\ \hline
  \end{tabular}
  \caption{Values of $\varphi_1$ and $\varphi_3$ for gCP transformations consistend with the residual Klein symmetry}
  \label{phitable}
  \end{table}
\end{center}
The neutrino mass matrix Eq.(\ref{mnu}) will be diagonalised by a unitary matrix $U_\nu$ via $U_\nu^T M_\nu U_\nu$. A matrix $U_\nu$ such that the diagonalised mass matrix is real and positive is given by
\begin{equation}
U_\nu^{(+)}=\left(
\begin{array}{ccc}
 -\frac{e^{i \left(-\frac{ \pi  \gamma }{n}-\frac{\varphi_1}{2} \right)}}{\sqrt{2}} & \frac{e^{ i \left(-\frac{ \pi 
   \gamma }{n}-\frac{\varphi_1}{2} \right)}}{\sqrt{2}} & 0 \\
 \frac{e^{i \left(\frac{\pi\gamma }{n}-\frac{\varphi_1 }{2}\right)}}{\sqrt{2}} & \frac{e^{i \left(\frac{\pi \gamma  }{n}-\frac{\varphi_1
   }{2}\right)}}{\sqrt{2}} & 0 \\
 0 & 0 & e^{-\frac{i \varphi_3}{2}} \\
\end{array}
\right)
\label{unuplus}
\end{equation}
for $|m_{21}|>|m_{22}|$ and  by
\begin{equation}
U_\nu^{(-)}=\left(
\begin{array}{ccc}
 -\frac{e^{ i \left(\frac{- \pi  \gamma }{n}-\frac{\varphi_1}{2}+\frac{\pi}{2} \right)}}{\sqrt{2}} & \frac{e^{ i \left(-\frac{ \pi 
   \gamma }{n}-\frac{\varphi_1}{2} \right)}}{\sqrt{2}} & 0 \\
 \frac{e^{i \left(\frac{\pi \gamma}{n}-\frac{\varphi_1}{2}+\frac{\pi}{2}\right)}}{\sqrt{2}} & \frac{e^{i \left(\frac{\pi \gamma }{n}-\frac{\varphi_1
   }{2}\right)}}{\sqrt{2}} & 0 \\
 0 & 0 & e^{-\frac{i \varphi_3}{2}} \\
\end{array}
\right)
\label{unuminus}
\end{equation}
for $|m_{21}|<|m_{22}|$. 

For charged leptons, the allowed gCP transformations with $X_r\in e^{i\alpha}G$ have to be consistent with $G_e=\{1,a,a^2\}$ and are given by
\begin{equation}
X_r=c^yd^{-y},ac^yd^{-y},a^2c^yd^{-y},bc^yd^{-y},abc^yd^{-y},a^2c^yd^{-y}
\end{equation}
where $3y=0 \text{ mod } n$. Especially when 3 divides $n$ there is a huge number of allowed $X$ matrices.
But, as the charged lepton mass matrix is already invariant under transformations with $a$ and transformations with $c^yd^{-y}$ force it to be zero (for $3y\neq 0 \text{ mod }n$) or produce no new constraint (for $3y=0\text{ mod }n$), the only transformations that produce physical constraints are given by
\begin{equation}
 X_r=\rho_r(1),\rho_r(b).
\end{equation}
For $X_r=\rho_r(1)$ the mass matrix of charged leptons is restrained to
\begin{equation}
M_{l1}M_{l1}^\dagger=\begin{pmatrix}
        m^e_3&m^e_1&m^e_2\\
        m^e_2&m^e_3&m^e_1\\
        m^e_1&m^e_2&m^e_3
       \end{pmatrix}
\end{equation}
with all parameters being real or for $X_r=\rho_r(b)$ to
\begin{equation}
M_{lb}M_{lb}^\dagger=\begin{pmatrix}
        m^e_3&m^e_1&(m^e_1)^\ast\\
        (m^e_1)^\ast&m^e_3&m^e_1\\
        m^e_1&(m^e_1)^\ast&m^e_3
       \end{pmatrix}
\end{equation}
with $m^e_1$ complex and $m^e_3$ real.
Both charged lepton mass matrices can be diagonalised by
\begin{equation}
U^e=\frac{1}{\sqrt{3}}
\begin{pmatrix}
1&1&1\\
\omega&\omega^2&1\\
\omega^2&\omega&1
\end{pmatrix}.
\end{equation}
Above charged lepton mass matrices only differ by unphysical phases which can be absorbed into the charged lepton fields. 

After removing an overall phase $e^{-i \varphi_1/2}$ to render the top left entry real, the physical mixing matrix is given by $U_\text{PMNS}^{(+)/(-)}=(U_e)^\dagger U_\nu^{(+)/(-)}$ (For $U_\nu^{(+)}$ and $U_\nu^{(-)}$ c.f. Eq.(\ref{unuplus}) and Eq.(\ref{unuminus})):
\begin{equation}
U_\text{PMNS}^{(+)/[(-)]}=
\left(
\begin{array}{ccc}
 \sqrt{\frac{2}{3}} \cos \left(\frac{\pi  \gamma }{n}\right) & \frac{e^{i(\varphi_1-\varphi_3)/2}}{\sqrt{3}} & [i]i
   \sqrt{\frac{2}{3}} \sin \left(\frac{\pi  \gamma }{n}\right) \\
 -\sqrt{\frac{2}{3}} \sin \left(\pi  \left(\frac{1}{6}+\frac{\gamma }{n}\right)\right) & \frac{e^{i(\varphi_1-\varphi_3)/2}}{\sqrt{3}} & [i]i \sqrt{\frac{2}{3}} \cos \left(\pi  \left(\frac{1}{6}+\frac{\gamma }{n}\right)\right) \\
 \sqrt{\frac{2}{3}} \sin \left(\pi  \left(\frac{1}{6}-\frac{\gamma }{n}\right)\right) & -\frac{e^{i(\varphi_1-\varphi_3)/2}}{\sqrt{3}} & [i]i \sqrt{\frac{2}{3}} \cos \left(\pi  \left(\frac{1}{6}-\frac{\gamma }{n}\right)\right) \\
\end{array}
\right)
\label{upmns}
\end{equation}
where the factor $i$ on the last column only appears in $U_\text{PMNS}^{(-)}$.
As the ordering of the mixing matrix is arbitrary at this point, we would like to fix it by requiring that the smallest entry of the matrix has to be the top-right entry, i.e. $U_{13}$. 
For small $\gamma/n$ the first row and third column are in the right place in the above matrix. 

As this matrix is now in the PDG convention, the values of Majorana phases $\alpha_{21}$ and $\alpha_{31}$ as well as the Dirac CP phase $\delta_{CP}$ for this ordering of the mixing matrix can be read off the matrix. 
Recall that the PDG convention is 
$
U_{\mathrm{PMNS}} = R_{23} U_{13} R_{12} P
$ 
in terms of $s_{ij}=\sin (\theta_{ij})$,
$c_{ij}=\cos(\theta_{ij})$, the Dirac CP violating phase $\delta_{CP}$ and
further Majorana phases contained in $P={\rm diag}(1,e^{i\frac{\alpha_{21}}{2}},e^{i\frac{\alpha_{31}}{2}})$.

The Majorana phase $\alpha_{21}$ is then given by
\begin{equation}
 \alpha_{21}=\varphi_1-\varphi_3
\end{equation}
With table [\ref{phitable}] follows that 
\begin{equation}
\varphi_1-\varphi_3=-\frac{6\pi(\gamma+x)}{n} \text{ for } X=c^{x}d^{2x+2\gamma},abc^xd^{2x}
\label{deltaphi1}
\end{equation}
or
\begin{equation}
\varphi_1-\varphi_3=-\frac{3\pi}{2}-\frac{6\pi(\gamma+x)}{n} \text{ for } X=c^{x}d^{2x+2\gamma+n/2},abc^xd^{2x+n/2}.
\label{deltaphi2}
\end{equation}
The values of all CP phases depend on the ordering of Eq.(\ref{upmns}) which needs to be changed for higher values of $\gamma/n$. The possible values of the CP phases can be found in table (\ref{phasesordering}). There, $U'$ denotes the mixing matrix after reordering such that the entry with the smallest absolute value is in the top right corner. As for every $\gamma/n$ the second and third row can be interchanged, which results in changing $\delta_{CP}$ by $\pi$ while changing the prediction for $U_{23}$ and $U_{33}$ and thus the prediction for $\theta_{23}$. 
The Dirac CP phase is hence predicted to be $0$ or $\pi$, and since the lepton mixing matrix has the tri-maximal form for the second column, referred to as TM2, this leads to the mixing sum rules 
$\theta_{23}=45^\circ \mp \theta_{13}/\sqrt{2}$ for $\delta_{CP}=0,\pi$, respectively,
as previously noted in \cite{KingVNA} (for a review of sum rules see \cite{King:2013eh}).

The prediction of $\alpha_{31}$ also depends on the order of these rows. In the table (\ref{phasesordering}) the second row of the mixing matrix after reordering it is indicated in the column $U_{23}'$. Improved measurements of $\theta_{23}$ will constrain this freedom of interchanging the second and third row.
\begin{center}
\begin{table}[h]
  \begin{tabular}{|l||l|l|l|l|l|l|l|l|}
    \hline
    $\gamma/n$&$U_{13}'$&$U_{23}'$&$\delta_{CP}^{(-)/(+)}$&$\alpha_{21}^{(-)}$&$\alpha_{21}^{(+)}$&$\alpha_{31}^{(-)}$&$\alpha_{31}^{(+)}$\\ \hline\hline
    
   $0/12\ldots1/12$ &$U_{13}$&$U_{23}$&0&$\varphi_1-\varphi_3$&$\varphi_1-\varphi_3$&$2\pi$&$-\pi$\\\hline
   
       		        &$U_{13}$&$U_{33}$&$-\pi$&$\varphi_1-\varphi_3$&$\varphi_1-\varphi_3$&$0$&$\pi$\\\hline
   
$1/12\ldots2/12$ &$U_{31}$&$U_{21}$&0&$\varphi_1-\varphi_3$&$\varphi_1-\varphi_3-\pi$&$2\pi$&$\pi$\\\hline
   
			&$U_{31}$&$U_{11}$&$-\pi$&$\varphi_1-\varphi_3$&$\varphi_1-\varphi_3-\pi$&$0$&$-\pi$\\\hline
   
$2/12\ldots3/12$&$U_{31}$&$U_{11}$&$0$&$\varphi_1-\varphi_3$&$\varphi_1-\varphi_3-\pi$&$0$&$-\pi$\\\hline
    
		        &$U_{31}$&$U_{21}$&$-\pi$&$\varphi_1-\varphi_3$&$\varphi_1-\varphi_3-\pi$&$2\pi$&$\pi$\\\hline
    
$3/12\ldots4/12$&$U_{23}$&$U_{13}$&$0$&$\varphi_1-\varphi_3+2\pi$&$\varphi_1-\varphi_3+2\pi$&$0$&$\pi$\\\hline
    
			&$U_{23}$&$U_{33}$&$-\pi$&$\varphi_1-\varphi_3+2\pi$&$\varphi_1-\varphi_3+2\pi$&$2\pi$&$-\pi$\\\hline
    
$4/12\ldots5/12$&$U_{23}$&$U_{33}$&$0$&$\varphi_1-\varphi_3+2\pi$&$\varphi_1-\varphi_3+2\pi$&$2\pi$&$-\pi$\\\hline
    
			&$U_{23}$&$U_{13}$&$-\pi$&$\varphi_1-\varphi_3+2\pi$&$\varphi_1-\varphi_3+2\pi$&$0$&$\pi$\\\hline
    
$5/12\ldots6/12$&$U_{11}$&$U_{31}$&$0$&$\varphi_1-\varphi_3+2\pi$&$\varphi_1-\varphi_3+\pi$&$2\pi$&$\pi$\\\hline
    
			&$U_{11}$&$U_{21}$&$-\pi$&$\varphi_1-\varphi_3+2\pi$&$\varphi_1-\varphi_3+\pi$&$0$&$-\pi$\\\hline
  \end{tabular}
  \caption{Values of CP phases after reordering for different values of $\gamma/n$ in $U_\text{PMNS}^{(-)/[(+)]}$. In each row, $\gamma/n$ can take arbitrary values in the interval indicated. $U'$ denotes the matrix after reordering.}
  \label{phasesordering}
  \end{table}
\end{center}
The key observable for Majorana phases is neutrino-less double beta decay ($0\nu\beta\beta$).
The effective mass of neutrinoless double-beta decay is given by
\begin{equation}
|m_{ee}|=|\frac{2}{3}m_1 \cos^2(\frac{\pi\gamma'}{n})+\frac{1}{3}m_2e^{i\alpha_{21}}+\frac{2}{3}m_3\sin^2(\frac{\pi\gamma'}{n})e^{i(\alpha_{31}-2\delta)}|
\end{equation}
with
\begin{equation}
m_1=m_l\text{ , }m_2=\sqrt{m_l^2+\Delta m_{21}^2}\text{ , }m_3=\sqrt{m_l^2+\Delta m_{31}^2}
\end{equation}
for normal ordering
and
\begin{equation}
m_1=\sqrt{m_l^2+\Delta m_{31}^2}\text{ , }m_2=\sqrt{m_l^2+\Delta m_{21}^2+\Delta m_{31}^2}\text{ , }m_3=m_l
\end{equation}
for inverted ordering, where $m_l$ is the mass of the lightest neutrino and 
\begin{equation}
\gamma'=\gamma\text{ mod } \frac{1}{6}.
\end{equation} 
The absolute values of the entries of the mixing matrix after reordering are periodic in $\gamma/n$ which is why one can simplify the analysis by defining $\gamma'$ in this way.

There are 8 cases to distinguish for combinations of phases. Adding a multiple of $2\pi$ will not change the effect of $\alpha_{21}$ or $\alpha_{31}-2\delta$. For this reason, for both Eq.(\ref{deltaphi1}) and Eq.(\ref{deltaphi2}) the 12 cases in table (\ref{phasesordering}) reduce to 8 cases of values for 
\begin{equation}
\bar\alpha_{21}=\alpha_{21}+6\pi\frac{\gamma+x}{n}\text{ , }\bar\alpha_{31}=\alpha_{31}-2\delta
\end{equation}
that are given by 
\begin{equation}
(\bar\alpha_{21},\bar\alpha_{31})=(0,0),(\pi/2,0),(\pi,0),(3\pi/2,0),(0,\pi),(\pi/2,\pi),(\pi,\pi),(3\pi/2,\pi).
\end{equation}
The by far most stringent constraint on $\gamma/n$ comes from the measurement of $\theta_{13}$. The current 3 sigma range for $\theta_{13}$ from \cite{CapozziCSA} yields values of $\gamma'/n$ in the range $0.0460\ldots 0.0627$.

It is generally fine to not only consider $\gamma'/n$ in this range but even $\gamma/n$ because changing $\gamma$ by $1/6$ only changes $\alpha_{21}$ by $\pi$, which is included in the four cases discussed above.

\begin{figure}
\raggedleft
\begin{subfigure}{0.49\linewidth}
\includegraphics[width=\linewidth]{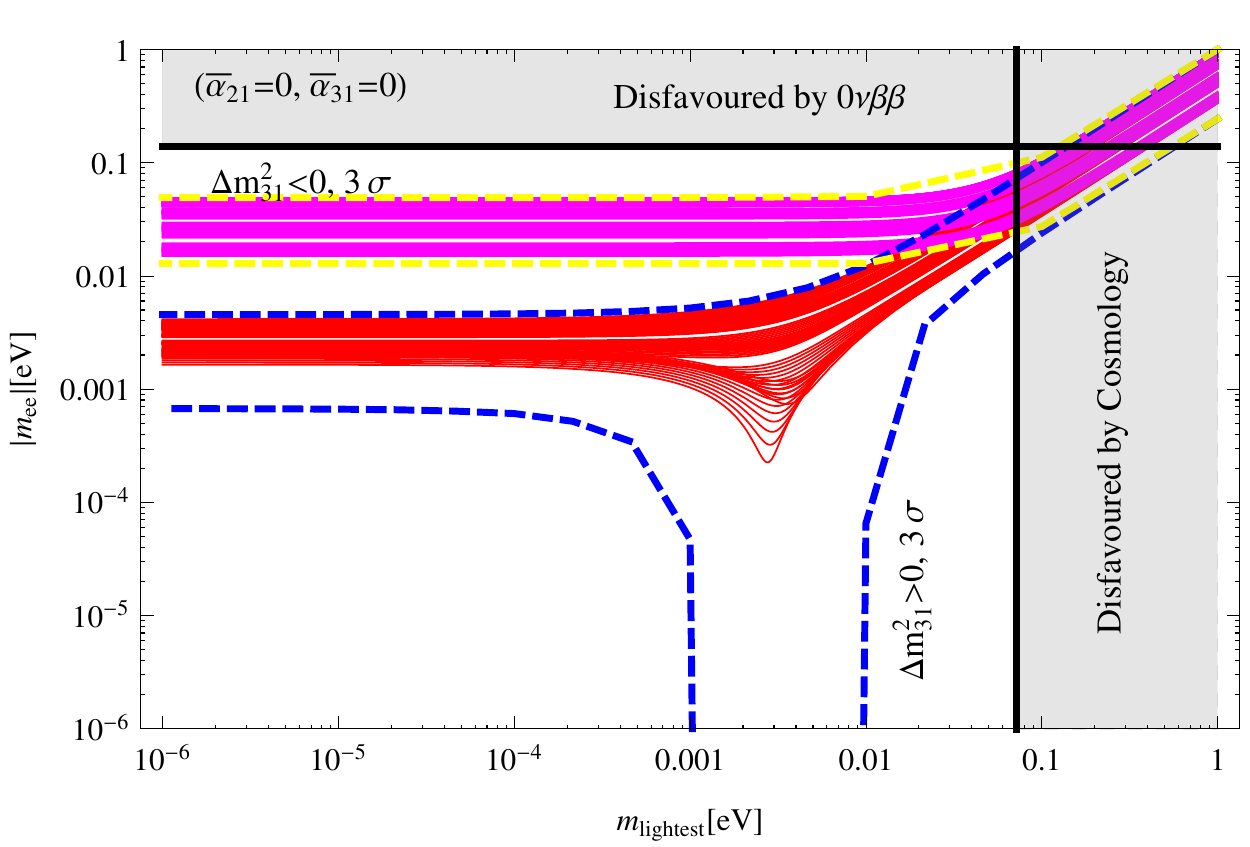}
\end{subfigure}
\begin{subfigure}{0.49\linewidth}
\includegraphics[width=\linewidth]{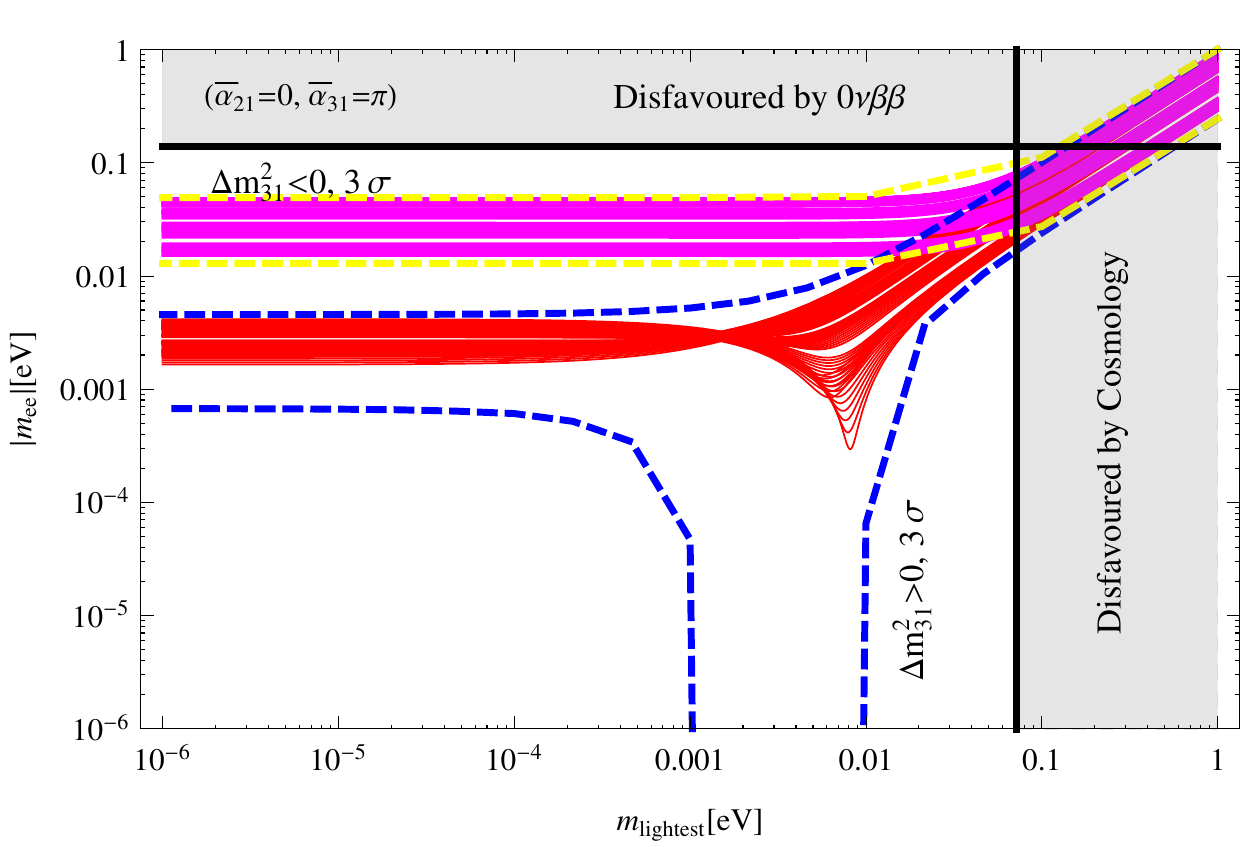}
\end{subfigure}
\begin{subfigure}{0.49\linewidth}
\includegraphics[width=\linewidth]{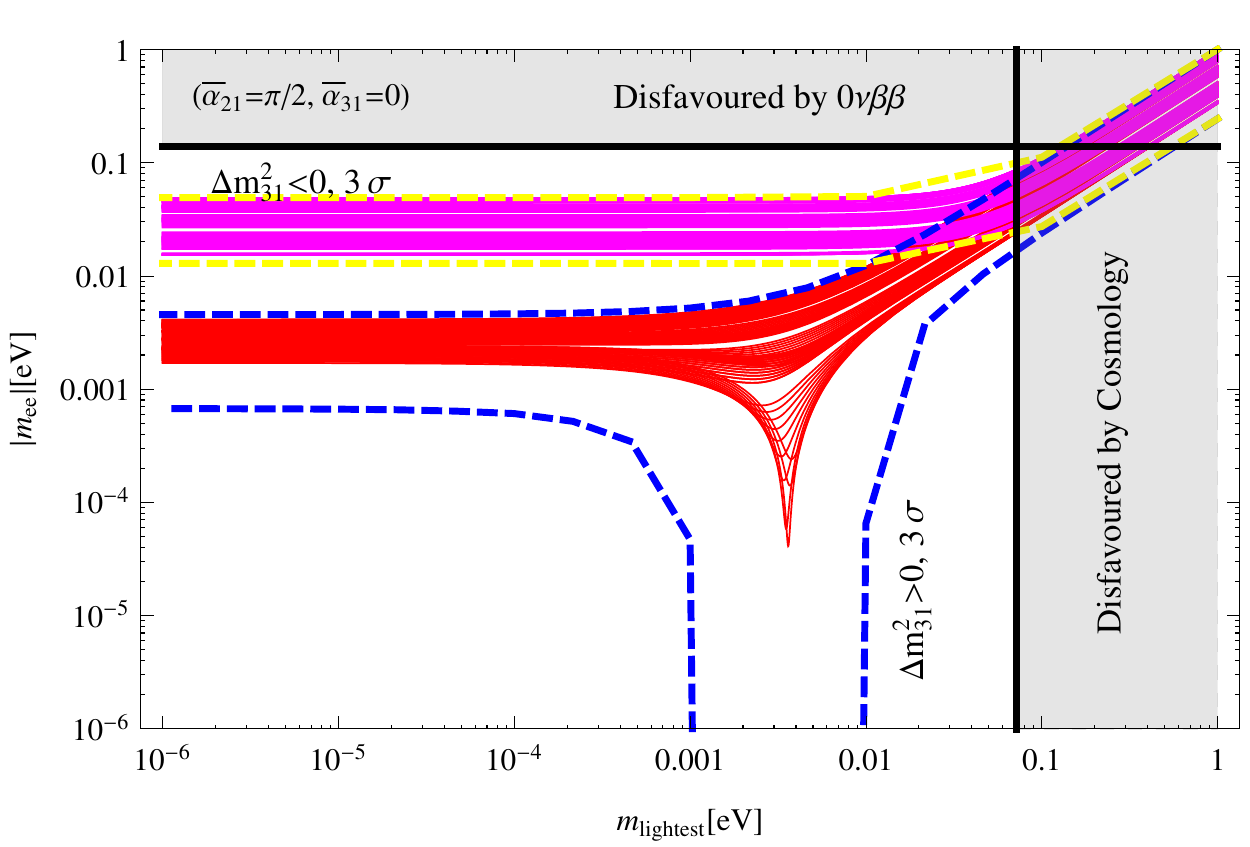}
\end{subfigure}
\begin{subfigure}{0.49\linewidth}
\includegraphics[width=\linewidth]{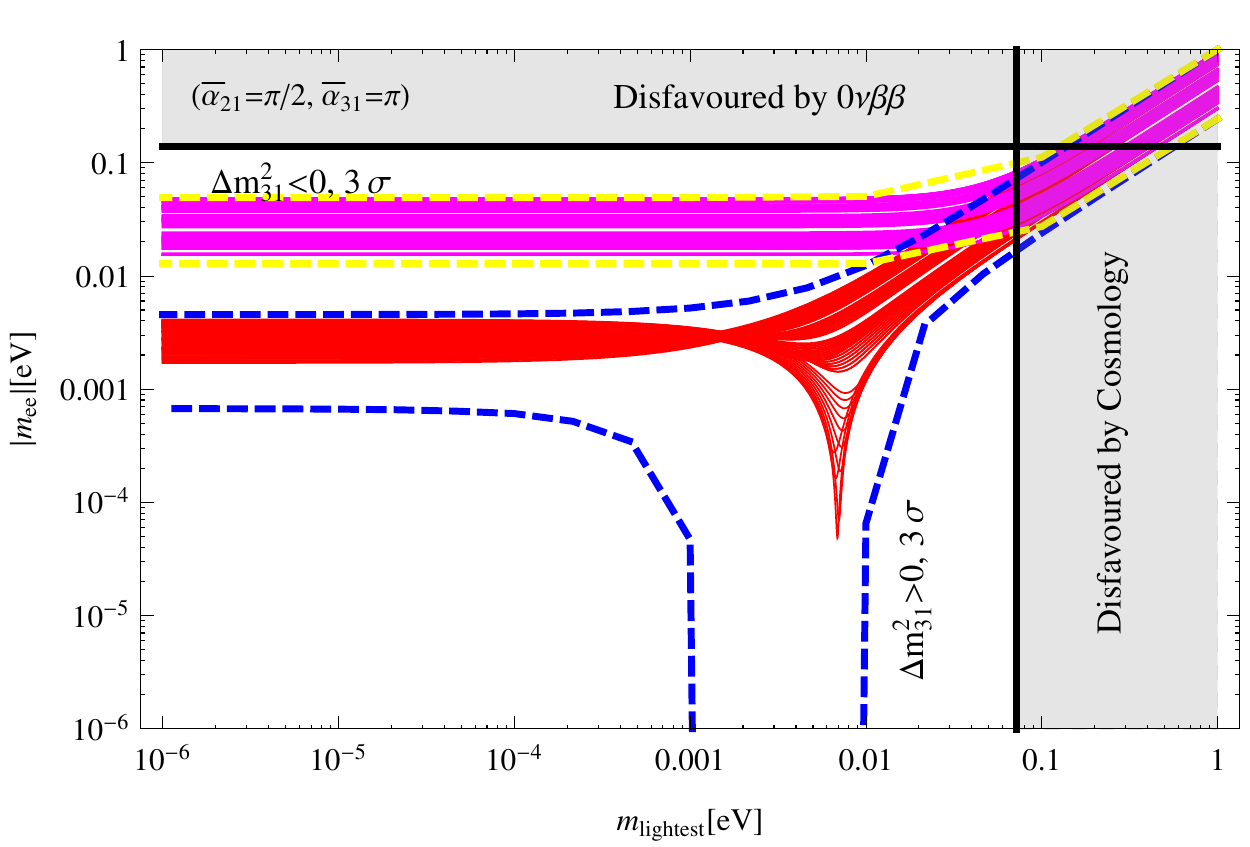}
\end{subfigure}
\begin{subfigure}{0.49\linewidth}
\includegraphics[width=\linewidth]{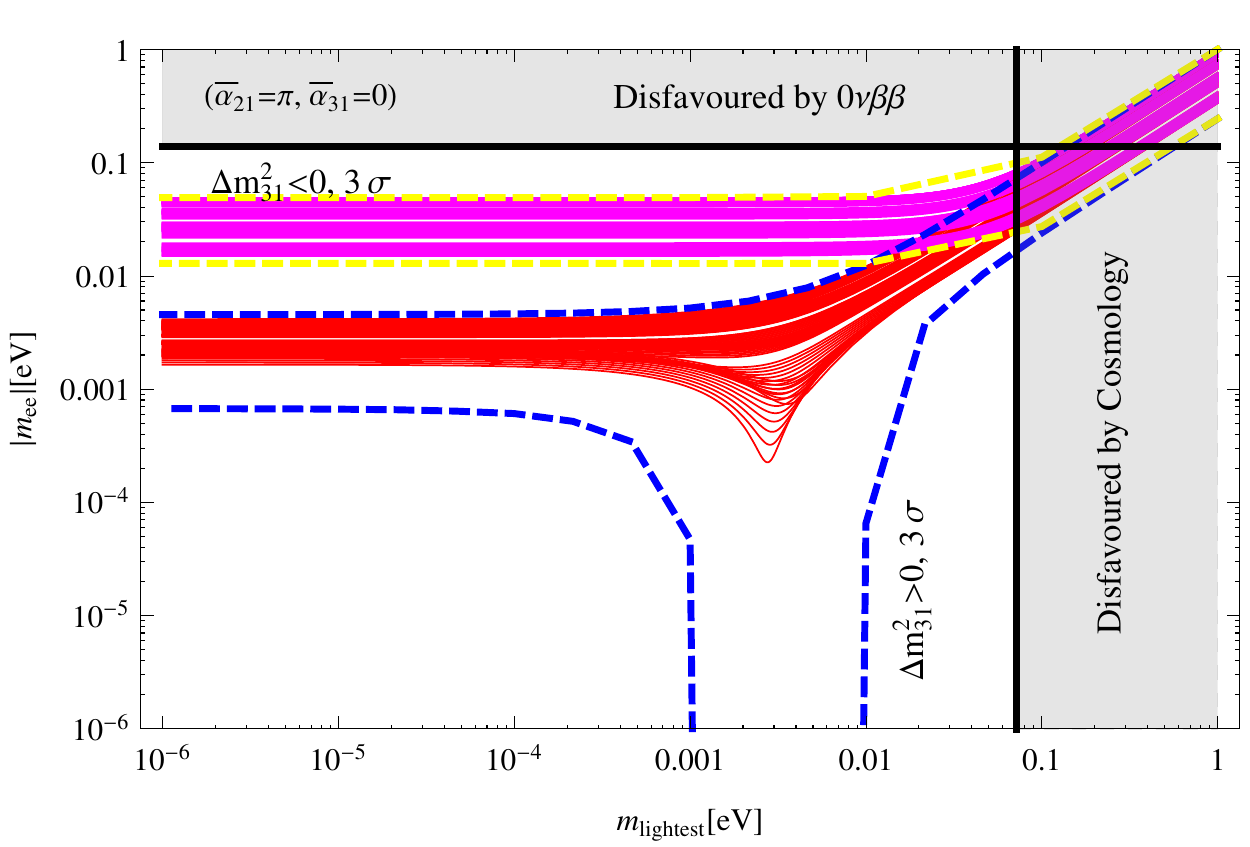}
\end{subfigure}
\begin{subfigure}{0.49\linewidth}
\includegraphics[width=\linewidth]{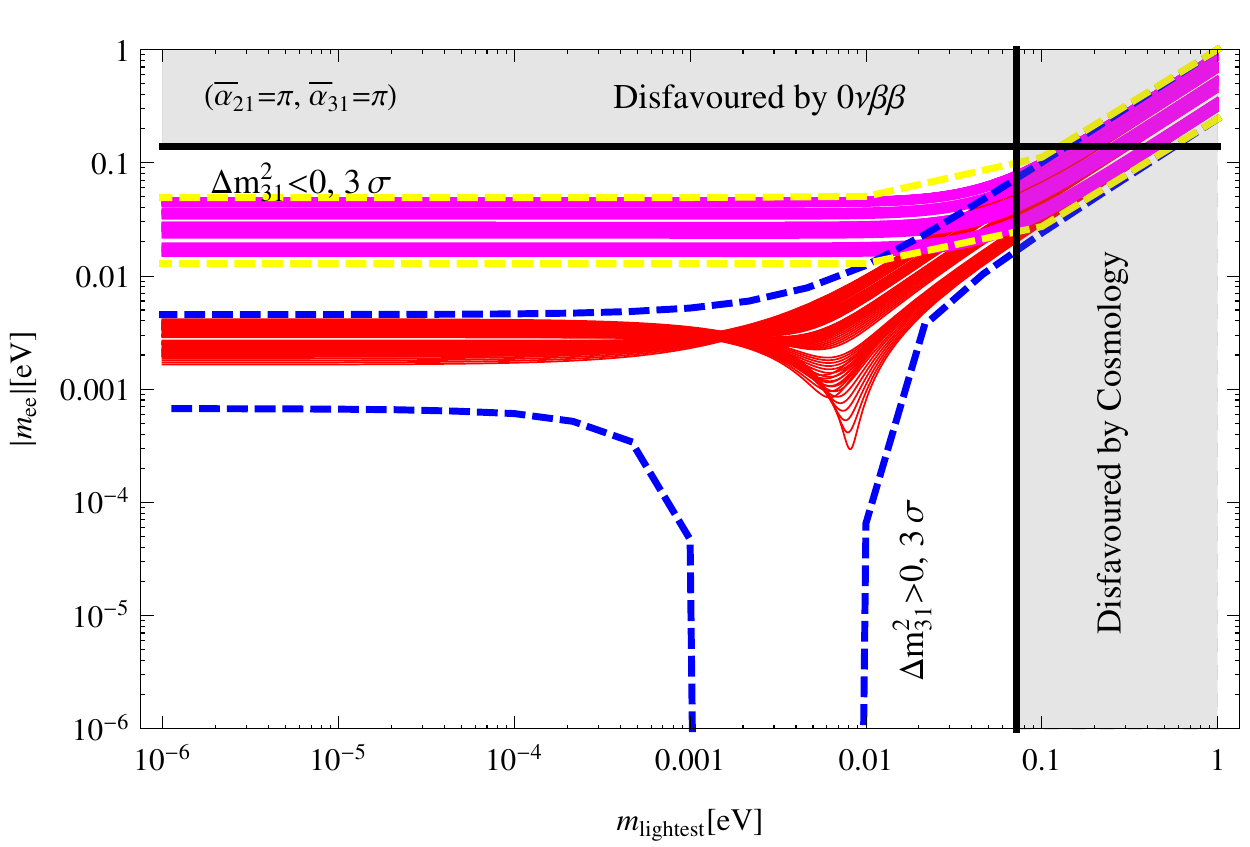}
\end{subfigure}
\begin{subfigure}{0.49\linewidth}
\includegraphics[width=\linewidth]{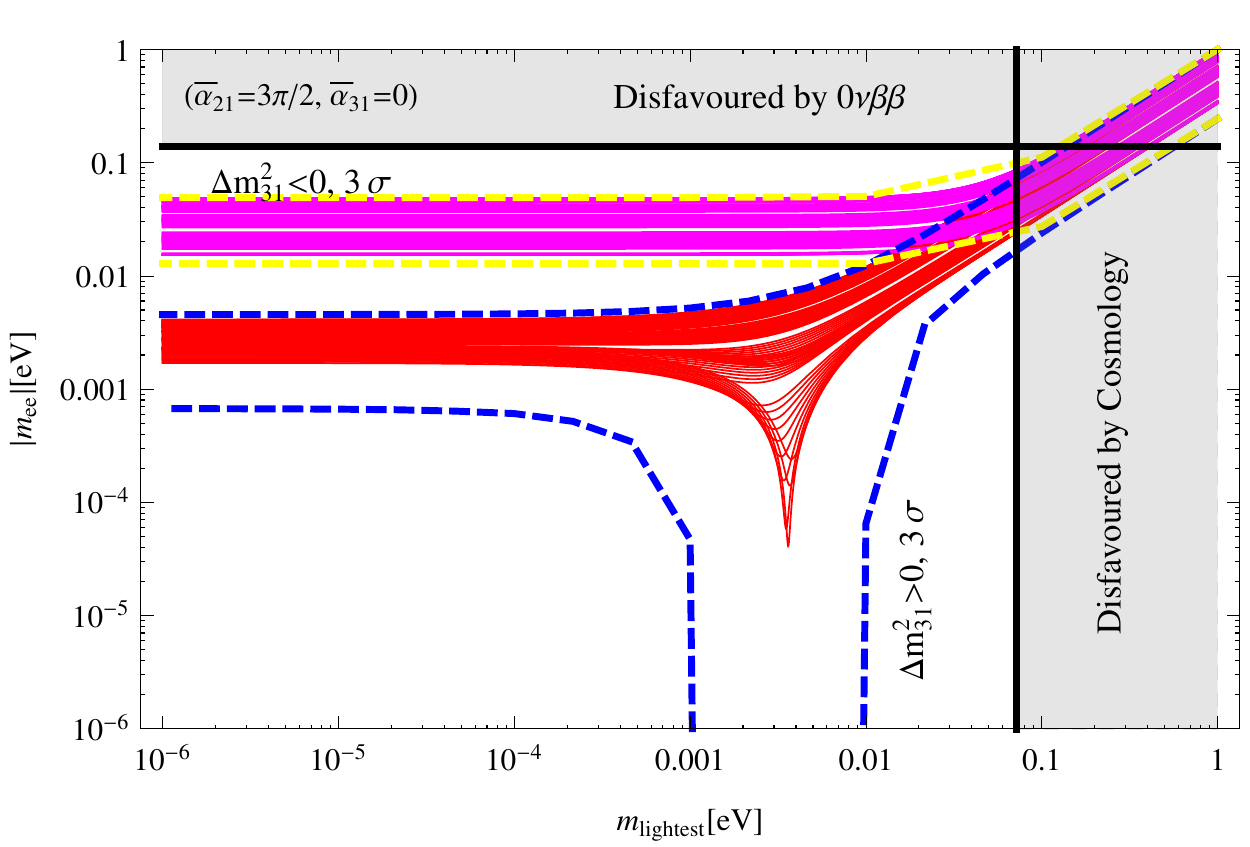}
\end{subfigure}
\begin{subfigure}{0.49\linewidth}
\includegraphics[width=\linewidth]{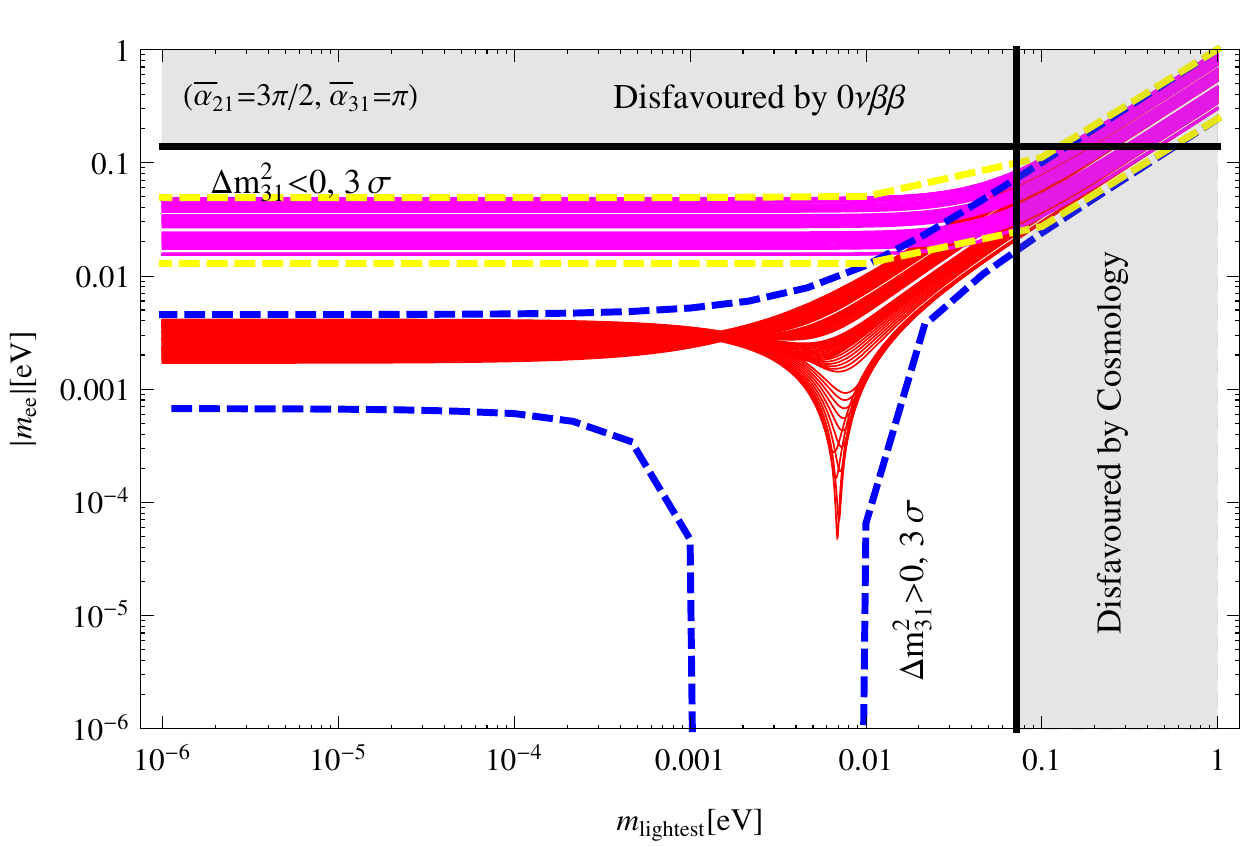}
\end{subfigure}
\caption{Effective Mass of $0\nu\beta\beta$ decay. $\gamma/n$ is varied between the lower and upper 3 sigma bound, $x/n=0,0.1,0.2,\ldots,1$. For the definition of $\bar\alpha_{21}$ and $\bar\alpha_{31}$ c.f. Eqs.(\ref{deltaphi1}), (\ref{deltaphi2}).}
\label{doublebeta}
\end{figure}

In order to understand predictions of $\Delta(6n^2)$ groups for $0\nu\beta\beta$ decay on a general level, in figure (\ref{doublebeta}) the effective mass $|m_{ee}|$ of $0\nu\beta\beta$ is plotted against the mass of the lightest neutrino $m_l$ for all combinations of $\bar\alpha_{21}$ and $\bar\alpha_{31}$. 
In these plots, models defined by some values of $\gamma/n$ and $x/n$ correspond to single fine lines. $\gamma/n$ takes 11 values, starting with the 3 sigma lower bound and increases in 10 equal steps until it reaches the 3 sigma upper bound. $x/n$ takes values $0,0.1,0.2,\ldots,1$. 

% Furthermore, each of the plots corresponds to one choice for the value of  $\alpha_{21}+6\pi\frac{\gamma+x}{n}$ and $\alpha_{31}-2\delta$, c.f. the captions of the plots. 
$\Delta m_{21}^2$ and $\Delta m_{31}^2$ are not varied, as doing so only would almost unnoticeably broaden each single line. Instead we used the best fit value from \cite{CapozziCSA}:
\begin{equation}
\Delta m_{21}^2=7.54\times 10^{-5} \text{ eV}^2,
\end{equation}
\begin{equation}
\Delta m_{31}^2=2.41\times 10^{-3} \text{ eV}^2.
\end{equation}

In figure (\ref{doublebeta}), Magenta lines correspond to predictions assuming inverted hierarchy, red lines to normal hierarchy.
Dashed blue and yellow lines indicate the currently allowed three sigma region for normal and inverted hierarchy, respectively. The three sigma ranges for mixing angles are taken from \cite{CapozziCSA}. The upper bound $|m_{ee}|<0.140$ eV is given from measurements by the EXO-200 experiment \cite{AugerAR}. Planck data in combination with other CMB and BAO measurements \cite{AdeZUV} provides a limit on the sum of neutrino masses of $m_1+m_2+m_3<0.230$ eV from which the upper limit on the mass of the lightes neutrino can be derived.

% 
% \begin{figure}
% \includegraphics[height=0.4\textheight]{doublebeta00new.pdf}
% \caption{Effective Mass of $0\nu\beta\beta$ decay for $\alpha_{21}+6\pi\frac{\gamma+x}{n}=0$ and $\alpha_{31}-2\delta=0$. $\gamma/n$ is varied between the lower and upper 3 sigma bound, $x/n=0,0.1,0.2,\ldots,1$.}
% \label{doublebeta00}
% \end{figure}
% \begin{figure}
% \includegraphics[height=0.4\textheight]{doublebeta01new.pdf}
% \caption{Effective Mass of $0\nu\beta\beta$ decay for $\alpha_{21}+6\pi\frac{\gamma+x}{n}=0$ and $\alpha_{31}-2\delta=\pi$. $\gamma/n$ is varied between the lower and upper 3 sigma bound, $x/n=0,0.1,0.2,\ldots,1$.}
% \label{doublebeta01}
% \end{figure}
% \begin{figure}
% \includegraphics[height=0.4\textheight]{doublebeta10new.pdf}
% \caption{Effective Mass of $0\nu\beta\beta$ decay for $\alpha_{21}+6\pi\frac{\gamma+x}{n}=\pi$ and $\alpha_{31}-2\delta=0$. $\gamma/n$ is varied between the lower and upper 3 sigma bound, $x/n=0,0.1,0.2,\ldots,1$.}
% \label{doublebeta10}
% \end{figure}
% \begin{figure}
% \includegraphics[height=0.4\textheight]{doublebeta11new.pdf}
% \caption{Effective Mass of $0\nu\beta\beta$ decay for $\alpha_{21}+6\pi\frac{\gamma+x}{n}=\pi$ and $\alpha_{31}-2\delta=\pi$. $\gamma/n$ is varied between the lower and upper 3 sigma bound, $x/n=0,0.1,0.2,\ldots,1$.}
% \label{doublebeta11}
% \end{figure}

The main features of the results from figure (\ref{doublebeta}) are as follows:

\begin{itemize}
\item 
For inverted hierarchy there is no particular structure visible. Additionally, the predicted values for $|m_{ee}|$ are well within the reach of e.g. phase III of the GERDA experiment of $|m_{ee}^\text{exp}|\sim0.02\ldots0.03$ eV \cite{KingPSA}.

\item
For normal ordering, it follows from figure (\ref{doublebeta}) that for the values of $\gamma/n$ and $x/n$ considered is always a lower limit on $|m_{ee}|$ which means that these parameters are accessible to future experiments.

\item
Further for normal ordering, in the very low $m_\text{lightest}$ region, predicted values of $|m_{ee}|$ are closer to the upper end of the blue three sigma range.

\item
With the current data, no combination of $\bar \alpha_{21}$ and $\bar \alpha_{31}$ is favoured. Only for values of $|m_{ee}|\lesssim0.0001$ eV and $m_\text{lightest}\lesssim0.01\ldots0.001$ eV it would be possible to distinguish different values of $\bar \alpha_{21}$ and $\bar \alpha_{31}$.

\end{itemize}

The necessary precisions on $|m_{ee}|$ and $m_\text{lightest}$ are unfortunately outside of the range of any projected experiments known to the authors. Nevertheless, the red curves corresponding to fixed values of $\gamma/n$ and $x/n$ are often close to the blue dashed three sigma range. With increasingly precise knowledge of the values of the mixing angles, especially $\theta_{13}$, the three sigma ranges will shrink, perhaps making it possible to draw conclusions about $\gamma/n$ and $x/n$ without an overly precise measurement of $|m_{ee}|$ or of the mass of the lightes neutrino.

To recapitulate, the following assumptions went into producing these results: There are 3 left-handed doublets of leptons, which in turn transform as a triplet under a $\Delta(6n^2)$ group. The neutrinos are Majorana fermions and $\Delta(6n^2)$ is broken to a $Z_2\times Z_2$ subgroup in the neutrino sector and to $Z_3$ in the charged lepton sector. The mixing angles are solely predicted from the aforementioned assumptions. There is a generalised CP symmetry which is consistent with $\Delta(6n^2)$ which is broken to one element in each sector. From this gCP symmetry the Majorana phases are predicted.

If one of the mixing angles would be found to be incompatible with any of the predictions this would mean that either $\Delta(6n^2)$ is not broken to $Z_2\times Z_2$ (or to $Z_2$, as the predictions for the mixing angles would be the same) or that the flavour group is not $\Delta(6n^2)$ or that one of the more fundamental assumptions is wrong. The neutrinos could still be Majorana fermions as $\Delta(6n^2)$ could still be broken completely.

\section{Conclusions}
In this paper we have examined the interplay of $\Delta(6n^2)$ groups and generalised CP transformations (gCP) in a direct model for three generations of Dirac charged leptons and Majorana neutrinos. We find that gCP transformations that actually are physical CP transformations have $X_r\in e^{i\alpha}\Delta(6n^2)$. Leaving a single gCP transformation unbroken will constrain the mixing matrix such that all phases, Dirac and Majorana are predicted and depend only on the $\Delta(6n^2)$ group, the residual $Z_2\times Z_2$ group (parametrised by $\gamma$) and the residual gCP transformation (parametrised by $x$) in the neutrino sector. Leaving two or more gCP transformations unbroken is not physically viable. 

Comparing the predictions for the mixing angles with experimental data we find that the strongest constraint on $\gamma/n$ is imposed by the relatively precise measurement of $\theta_{13}$. The smallest group where $\theta_{13}$ lies within three sigma of the central value has $n=14$.
Furthermore, since the Majorana CP violating phases are predicted,
we have studied predictions for neutrinoless double-beta decay. We find that for inverted ordering, the predicted $|m_{ee}|$ is within the reach of upcoming experiments like GERDA III. For normal ordering, measuring $|m_{ee}|$ down to $10^{-4}$eV could exclude large regions of $\gamma/n$ and $x/n$, depending on the value of $\delta_{CP}$.

In conclusion, this paper represents the first time that an infinite series of finite groups has been examined for generalised CP transformations that are consistent with it. We emphasise the important role of 
$\Delta(6n^2)$ among the subgroups of $SU(3)$ with triplet irreducible representations and hope that this study will help to shed some light on the mystery of neutrino mixing. 
If the Dirac CP phase is measured to differ from 0 or $\pi$, or the mixing angles deviate from the sum rules 
$\theta_{23}=45^\circ \mp \theta_{13}/\sqrt{2}$, respectively, then
this would mean that in general a potential flavour group $\Delta(6n^2)$ cannot be broken to $Z_2\times Z_2$, as in the case of the direct approach assumed here. 
However the semi-direct approach, in which 
a $Z_2$ subgroup is preserved, would remain a possibility for theories based on $\Delta(6n^2)$.

\section*{Acknowledgments}
The authors would like to thank Alex Stuart for many very useful discussions. The authors acknowledge partial support from the European Union FP7 ITN-INVISIBLES (Marie Curie Actions, PITN- GA-2011- 289442). SFK acknowledges support from the STFC Consolidated ST/J000396/1 grant.

\end{document}